# The *MAPT* gene is differentially methylated in the progressive supranuclear palsy brain

Vincent Huin[1*], MD; Vincent Deramecourt[1], MD, PhD; Dominique Caparros-Lefebvre[2], MD, PhD; Claude-Alain Maurage[1], MD, PhD; Charles Duyckaerts[3] MD, PhD; Eniko Kovari[4], MD; Florence Pasquier[5], MD, PhD; Valérie Buée-Scherrer[1], PhD; Julien Labreuche[3], BST; Hélène Behal[3], BST; Luc Buée[1], PhD; Claire-Marie Dhaenens[1], PharmD, PhD; Bernard Sablonnière[1], MD, PhD

1. Univ. Lille, Inserm, CHU Lille, UMR-S 1172 - JPArc - Centre de Recherche Jean-Pierre AUBERT Neurosciences et Cancer, F-59000 Lille, France

2. Centre Hospitalier de Wattrelos, Service de Neurologie, Wattrelos, France

3. Laboratoire de Neuropathologie Escourolle, AP-HP, Hôpital de la Pitié Salpêtrière, 75013, Paris, France

4. Department of Mental Health and Psychiatry, University Hospitals and University of Geneva, 1225, Geneva, Switzerland

5. Univ. Lille, Inserm, CHU Lille, U1171 - CNR-MAJ, DISTALZ, F-59000 Lille, France

6. Univ. Lille, CHU Lille, EA 2694 - Santé publique : épidémiologie et qualité des soins, Département de Statistiques, F-59000 Lille, France

*corresponding author: Vincent Huin

INSERM UMR-S 1172, Centre de Recherche Jean-Pierre Aubert, Bâtiment Biserte, 1, Place de Verdun, F-59045, Lille, France

vincent.huin@inserm.fr

+33 (0)3 20 62 20 75

**Running head:** Hypomethylation of *MAPT* in PSP brain

**Number of words in the entire manuscript (not including Title Page, Abstract, Figure Legends & References):** 3621 words

**Number of characters in the title:** 86 characters

**Number of characters in the running head:** 37 characters

**Number of words in the abstract:** 265 words

**Number of references:** 40

**Number of figures, color figures, and tables:** 4 figures (1 color figure), 1 table, 4 supplemental figures, 1 supplemental table

**ABSTRACT:**

**Background:** Progressive supranuclear palsy (PSP) is a rare neurodegenerative disease causing parkinsonian symptoms. Altered DNA methylation of the microtubule-associated protein tau gene correlates with the expression changes in Alzheimer's disease and Parkinson's disease brains. However, few studies examine the sequences beyond the constitutive promoter.
**Objectives**: Because activating different microtubule associated protein tau gene control regions via methylation might regulate the differential tau expression constituting the specific signatures of individual tauopathies, we compared methylation of a candidate promoter, intron 0.
**Methods:** We assessed DNA methylation in the brains of patients with different tauopathies (35 Alzheimer's disease, 10 corticobasal degeneration, and 18 PSP) and 19 controls by intron 0 pyrosequencing. We also evaluated methylation in an independent cohort of 11 PSP cases and 12 controls. Frontal (affected by tau pathology) and occipital (unaffected) cortices were analyzed.
**Results:** In the initial samples, one CpG island site in intron 0 (CpG1) showed significant hypomethylation in PSP-affected frontal cortices when compared with controls ($p = 0.022$). Such hypomethylation was observed in replicate samples, but not in occipital cortices or other tauopathies. PSP and control samples (combining the initial and replicate samples) remained significantly different after adjustment for potential confounding factors (age, H1/H1 diplotype; $p = 0.0005$). PSP-affected tissues exhibited microtubule-associated protein tau RNA hyperexpression when compared with controls ($p = 0.004$), although no correlation with CpG1 methylation was observed.
**Conclusions:** This exploratory study suggests that regions other than the constitutive promoter may be involved in microtubule-associated protein tau gene regulation in tauopathies and that intron 0 hypomethylation may be a specific epigenetic signature of PSP. These preliminary findings require confirmation.

**Conflicts of interest:** All authors report no financial disclosures or conflicts of interest related to the research in the article.

**Funding:** This study was supported by the LabEx Development of Innovative Strategies for a Transdisciplinary approach to ALZheimer's disease (DISTALZ) and the Association paralysie supranucléire progressive France.

**<u>Abbreviations:</u>**

AD: Alzheimer's disease

AP: Alternative promoter

CBD: Corticobasal degeneration

MAPT: Microtubule-associated protein tau

PD: Parkinson disease

PCR: Polymerase chain reaction

PMI: Post-mortem interval

PSP: Progressive supranuclear palsy

qRT-PCR: Quantitative reverse transcription-polymerase chain reaction

SD: Standard deviation

SEM: Standard error of the mean

SNP: Single nucleotide polymorphism

TALE: Transcription activator-like effector

TSS: Transcription start site

# INTRODUCTION:

Progressive supranuclear palsy (PSP), or Steele–Richardson–Olszewski syndrome, is a rare neurodegenerative disease characterized clinically by vertical supranuclear gaze palsy, postural instability, bradykinesia, behavioral modifications, and cognitive decline.[1-3] This disorder belongs to the tauopathies, a group of neurodegenerative diseases defined by abnormal aggregation of hyperphosphorylated tau protein in the brain. In PSP, tau accumulates within neurons as neurofibrillary tangles and in glial cells as tufted astrocytes.[4] This tau pathology also has a specific pattern of distribution and progression in the brain that correlates with the clinical course. The stereotypical progression of tau accumulation in PSP typically starts in the basal ganglia and motor cortex. Lesions then progress to the brain stem, the entire frontal cortex, and the cerebellum.[5,6] As with many other tauopathies, PSP is more frequent in aged people, with a mean onset age of 63 years.[3] Environmental and/or lifestyle factors that potentially contribute to the risk of PSP remain unknown. A genome-wide association study performed on 1114 autopsied cases and 1051 living patients diagnosed with PSP identified common genetic variations that were highly associated with the disease ($p < 5.10^{-8}$) in the genes syntaxin 6 (*STX6*), eukaryotic translation initiation factor 2-alpha kinase 3 (*EIF2AK3*), myelin-associated oligodendrocyte basic protein (*MOBP*), and microtubule-associated protein tau (*MAPT*).[7] The association between *MAPT* and PSP was first described in 1999. Indeed, a common inversion of 900 kb encompassing a number of genes, including *MAPT*, defines the following 2 haplotypes: H1 and H2. The H1 haplotype is more frequent and is significantly overrepresented in multiple neurodegenerative diseases, including PSP[8] and corticobasal degeneration (CBD)[9] and likely to a lesser extent in Alzheimer's disease (AD) patients without the apolipoprotein E (*APOE*) ε4 genotype (odds ratio = 1.12; $p = 0.0005$).[10] In PSP, H1 appears to be the major risk factor, with an odds ratio of 5.46.[7] In the background of this H1 clade, the H1c sub-haplotype, defined by the A allele of the single-nucleotide polymorphism (SNP) rs242557 (located in intron 0 of *MAPT*), is highly associated with PSP[10,11] and with elevated expression of *MAPT*.[12] Despite the information accumulated to date, the pathophysiological mechanisms underlying these risk factors remain unclear.

Epigenetic changes in neurodegenerative diseases have recently begun to attract attention. Changes in DNA methylation are well known to correlate with aging.[13] Moreover, Li and collaborators[14] recently proposed that the increased risk of PSP as a result of the H1 haplotype could largely be explained via differential methylation at the *MAPT* locus. *MAPT* has previously been explored in AD[15] and Parkinson's disease (PD),[16] whereby altered DNA methylation of *MAPT* was associated with abnormal gene expression in patients' brains. These 2 studies assessed the DNA methylation of the main CpG island (CpGI) of *MAPT*, which encompasses the large, noncoding exon 0. However, few studies have examined genomic regions other than the *MAPT* constitutive promoter,[17] and it is now expected that DNA methylation at intragenic CpGIs may regulate the use of alternative intragenic promoters.[18] DNA methylation at these sites may increase chromatin accessibility to binding factors and transcription initiation factors, which may therefore affect transcription (for review, see Kulis et al.[19]. Our hypothesis is that DNA methylation-based activation of different control regions of *MAPT* may be a key factor in the differential tau expression that constitutes the specific signature of each tauopathy. To investigate a new *MAPT* regulatory sequence specific to PSP brains, we assessed DNA methylation of a candidate region in autopsied brains of PSP patients when compared with other tauopathies and controls. We searched for regions in *MAPT* that exhibit accumulation of epigenetic marks compatible with enhancer or promoter functions. We compared DNA methylation in frontal cortices with neurofibrillary degeneration to occipital samples without pathological tau from the same patients.

## **PATIENTS AND METHODS:**

### **Human brain tissue samples**

Brain tissue samples were obtained from the brain banks of clinical centers in Lille, Paris, and Geneva, according to the procedures of the local ethics committees. Dissection and diagnosis were conducted by trained neuropathologists and confirmed by Western blotting using phospho-specific tau antibodies to reveal pathological tau proteins. Initially, this study included 19 control individuals and 63 patients (35 AD, 10 corticobasal degeneration [CBD], and 18 PSP). To replicate the main DNA methylation results, an independent cohort of 12 controls and 11 PSP brains from a Paris clinical center was added. Controls were defined as individuals with no signs of cognitive decline, previous stroke, or chronic brain pathology, with Braak stages of 0 to 2. Patients with multiple neurodegenerative diseases were excluded. Subjects with a postmortem interval greater than 48 hours, age at death younger than 40 years, or a history of brain tumors, intracranial bleeding, or inflammatory brain disease were also excluded from the sample. Frontal and occipital cortices were dissected for analysis. The main clinical, neuropathological, and genetic features of the included subjects are described in Table 1.

### **DNA and RNA extraction**

We excised 200mg of gray matter from the frontal lobe (Brodmann area 10; $n = 105$) and from the occipital lobe (Brodmann area 18; $n = 96$). One sample of 50 mg was used for DNA extraction, and another 50 mg was used for RNA extraction. Genomic DNA was obtained by phenol-chloroform extraction. Total RNA was isolated from human brain samples by using the RNeasy Lipid Tissue Mini Kit (Qiagen, Courtaboeuf, France) according to the manufacturer's instructions. RNA quality was determined based on 28S/18S rRNA readings and on the RNA integrity number obtained from an Agilent Bioanalyzer 2100 using the RNA 6000 Nano Kit (Agilent, Courtaboeuf, France). Only samples with a concentration of >100 ng/ml and RNA integrity number ≥5 (*n* = 94) were retained for subsequent quantitative reverse transcription-polymerase chain reaction (qRT-PCR; Supplemental Fig. S1).

**Pyrosequencing**

All DNA samples and methylated genomic DNA controls from the Human Premixed Calibration Standard (EpigenDX, Hopkinton, Massachusetts) were subjected to sodium bisulfite treatment using the EZ DNA Methylation Kit (Zymo Research, Irvine, California) according to the manufacturer's specifications. Briefly, 500 ng of bisulfite-treated DNA was amplified by PCR. Quantitative methylation analyses were performed by pyrosequencing using the PyroMark MD system (Biotage, Uppsala, Sweden; Qiagen), and the results were analyzed with the Pyromark Q24 software (Qiagen). All CpG sites were tested in duplicate using 2 different bisulfite conversions.

**Genotyping**

The observed variations in methylation levels were examined and separated based on the H1 haplotype and the sub-haplotype H1c. The H1/H2 haplotype was characterized by the presence of the 238 bp insertion/deletion in *MAPT* intron 9.[8] Genotyping of SNP rs242557, which defines the H1c sub-haplotype, was performed using standard PCR and sequencing on an ABI3730 DNA analyzer (Applied Biosystems, Saint Aubin, France). PCR, sequencing, and pyrosequencing primers are listed in Supplemental Table S1.

**Quantitative RT-PCR**

For each of the 90 RNA samples, 1 μg of total RNA was used to generate cDNA using the High-Capacity cDNA Reverse Transcription Kit (Applied Biosystems™, Saint Aubin, France) with Multiscribe Reverse Transcriptase and random primers. *MAPT* expression levels were determined by qRT-PCR, using TaqMan® Gene Expression Assays, TaqMan® Universal PCR Master Mix II, with Uracil-N glycosylase (Life Technologies, San Francisco, CA, USA). TaqMan® assays were used to quantify mRNA levels of *MAPT* (Hs00902194_m1) using *UBC* (Hs00824723_m1) as the control housekeeping gene. All qRT-PCR were performed on an ABI PRISM® 7900 HT instrument (Applied Biosystems™) in triplicate for each sample, according to the manufacturer's protocol. The comparative CT method ($2^{-\Delta\Delta CT}$) was used to calculate the relative expression levels.

**Statistical analysis**

Major characteristics are described for each study group without statistical comparisons. Qualitative variables are expressed as the number (percentage) for each group. Continuous variables are expressed as the mean (± standard deviation [SD]) in cases of normal distribution and as the median (± interquartile range) otherwise. The normality of distributions was assessed using histograms and the Shapiro–Wilk test. After pooling all groups, we assessed the pairwise correlations of methylation levels (in frontal and occipital areas separately) at the 5 CpGs by calculating the Pearson correlation coefficients. Comparisons of the frontal and occipital methylation levels of the 5 CpGs between controls and each disease group were performed using analysis of variance with Dunnett's post hoc test. Using analysis of covariance, comparisons of methylation levels between PSP and controls were further adjusted for 2 potential confounding factors (age at death and H1/H1 diplotype) based on prior evidence of an association with PSP. For CpG1, for which frontal methylation levels differed significantly between controls and PSP, a replication analysis was performed using an independent cohort of 11 individuals with PSP and 12 controls; methylation levels were compared using Student's *t* test. Finally, given that hypomethylation is associated with upregulated gene expression, we compared mRNA expression between the controls and diseases using the Mann–Whitney U test. Statistical testing was performed at the 2-tailed $\alpha$ level of 0.05. Data were analyzed using the software package SAS, release 9.3 (SAS Institute, Cary, North Carolina).

## RESULTS:

The main characteristics of the 2 cohorts are described in Table 1, divided by study groups. Several clear differences between groups were observed, with a greater mean age for AD patients, different sex ratios for controls and tauopathy patients, and, as expected, an increased H1/H1 diplotype frequency in PSP patients.

### *MAPT* candidate region and CpGs selection

Epigenetic marks in the *MAPT* gene that could define candidate regions for regulating *MAPT* expression were selected using the GRCh37/hg19 build of the University of California, Santa Cruz (UCSC) Genome Browser (http://genome.ucsc.edu/). A region encompassing the second CpGI in intron 0 and its shores (located between g.44026528 and g.44026738 on chromosome 17) was chosen because it contains several traits representative of a regulatory region, such as a CpGI, clusters of DNase I, transcription factor, and CCCTC-binding factor (CTCF) binding sites and H3K27Ac (Fig. 1). All of these modifications are compatible with an enhancer or a weak promoter. Moreover, this area is located 7 kb downstream of rs242557, the SNP defining the H1c sub-haplotype. The methylation levels of 5 CpGs were assessed in this region. The first CpG, arbitrarily named CpG1, is located in the CpG shore upstream of the candidate sequence, whereas the other 4 are located in the CpG island.

### Hypomethylation in PSP brains

Methylation was assessed for all 5 CpGs in the 2 brain areas for the entire cohort. When all groups were pooled, the methylation levels of the 5 CpGs correlated positively with each other in the frontal area and in the occipital area. The strongest correlations were identified between CpGs 28 and 30 (all *r*>0.75, see Supplemental Fig. S2). No significant difference between controls and each tauopathy group was observed for the frontal methylation levels of CpGs 27 and 30 (all *p*>0.23, see Supplemental Fig. S3). Similarly, no significant difference was observed in occipital samples for the 5 CpGs. Interestingly, we identified a significant difference in the frontal methylation levels of CpG1 between the controls and PSP group. The mean (± SD) CpG1 methylation level in the frontal area was reduced in the PSP group (35.6 ± 2.8 vs 39.7 ± 4.1 in controls, *p* = 0.022), whereas no such difference was observed for CpG1 methylation in the occipital area (31.5 ± 2.9 in PSP vs 30.5 ± 3.1 in controls, *p* = 0.85; Fig. 2). Moreover, no significant differences were observed for CpG1 in the frontal or occipital areas between the controls and the AD or CBD groups. Thus, methylation analysis revealed DNA hypomethylation in a part of the targeted region of *MAPT* intron 0 only in brain tissue with neurofibrillary degeneration.

The difference in frontal methylation levels of CpG1 between PSP patients and controls remained significant in multivariate analyses adjusted for age at death and H1/H1 diplotype (adjusted means [± standard error of the mean]: 36.2 ± 0.9 vs 39.4 ± 0.8 in controls, *p* = 0.012), whereas the differences in all other comparisons (methylation levels in other CpGs) remained nonsignificant.

**Replication analysis of hypomethylation in PSP brains**

The replicate samples included 11 PSP brains (mean age at death, 74 ± 8; men, 45%; H1/H1 diplotype, 63.6%) and 12 controls (mean age at death, 79 ± 14; men, 50%; H1/H1 diplotype, 66.7%). As in the previous analysis, the CpG1 methylation level in the frontal area was significantly lower in PSP brains (mean ± SD, 33.1 ± 6.3) than in control brains (mean ± SD, 38.5 ± 4.6; *p* = 0.029). No difference was found in the level of CpG1 methylation in the occipital area (Fig. 3). When replicate samples were combined with initial samples, the difference in frontal methylation levels of CpG1 between PSP patients and controls was significant after adjusting for age at death and H1/H1 diplotype, with an adjusted mean difference (PSP vs controls) of -4.2 (95% confidence interval, -6.4 to -1.9; *p* = 0.0005).

**Hyperexpression of *MAPT* mRNA in PSP brains**

Hypomethylation is associated with an upregulation of gene expression. To verify this observation in our cohort, we measured mRNA expression in all samples by qRT-PCR. As shown in Figure 4, *MAPT* expression was significantly increased in PSP patients when compared with controls in the frontal area (*p* = 0.004), but not in the occipital area (p = 0.65). Specifically, *MAPT* expression in PSP-affected brain tissues was increased 1.6-fold when compared with controls. Although AD patients showed higher *MAPT* expression than controls, this difference did not reach significance for either area (p = 0.077 in the frontal and p = 0.071 in the occipital area). However, we found that DNA methylation of CpG1 did not correlate with *MAPT* expression in any of the groups (Supplemental Fig. S4).

**DISCUSSION:**

The aim of our exploratory study was to assess whether DNA methylation of different and poorly characterized regions of *MAPT* could affect the differential expression of *MAPT* in tauopathies. Our results indicate that methylation of a specific CpG called CpG1 in a novel control region within intron 0 of the *MAPT* gene is associated with PSP. Finally, hypomethylation of intron 0 was associated with increased *MAPT* mRNA expression in PSP-affected brain tissues.

Altered DNA methylation of the *MAPT* gene has previously been investigated in AD,[12,20] PD,[16] and other neurodegenerative diseases.[15-17,20] Studies on AD, the most prevalent cause of dementia worldwide, assessed *MAPT* methylation in several brain regions that are important in neuropathological processes, such as the hippocampus,[17,21] frontal cortex,[17] and temporal cortex.[15] However, considering the disease frequency, the various brain cohorts reported are relatively small. Sanchez-Mut and colleagues[21] compared 25 AD brains with 25 controls, and Iwata and colleagues compared as many as 59 AD brains with 76 controls.[15] Regarding the genomic region investigated, most previous studies examining the CpGI encompassed exon 0 and its promoter. Precisely, these studies focused on the core promoter, which extends from approximately 640 bp upstream of the transcription start site (TSS) in exon 0[17] to 211 bp downstream of the TSS[20] (Fig. 1). The number of CpGs investigated to date is rather limited (from 1 to 41),[16,17,21-23] encompassing regions of up to a few hundred base pairs, with significant results observed for 1 to 5 differentially methylated CpGs. For example, Coupland and colleagues16 targeted only 1 SNP (rs76594404) located in the core promoter (291 bp upstream of exon 0) to assess methylation in 28 PD brains versus 12 controls. Although Iwata and colleagues screened the main CpGI of *MAPT* in 59 AD patients and 76 controls, DNA hypomethylation was observed only in 5 consecutive CpGs located between 20 and 30 bp after the start of exon 0. Interestingly, Barrachina and colleagues[17] explored 31 CpGs 54,336 to 54,905 bp downstream of exon 0, in the same region as we investigated, but focused on the CpGI. Their study focused on 2 brain regions (the frontal cortex and hippocampus) from 98 patients who presented with various neurodegenerative diseases, including AD ($n = 44$) and Pick's disease ($n = 3$), compared with 26 controls. The authors found no significant results because of the low precision of the Matrix-assisted laser desorption/ionization-time-of-flight mass spectrometry (MALDI-TOF) technology used. Regarding PSP, the cohorts of postmortem brains described in the literature are relatively small because of the low prevalence of the disease, which is approximately 5 to 6 cases per 100,000 persons.[3,24,25] Indeed, larger cohorts are actually clinicopathological descriptions of as many as 100 cases in multicentric studies.[24-27] Only 1 recent study assessed genome-wide DNA methylation in 43 PSP brains (temporal cortex) from 175 PSP cases and compared them with 185 controls.[28] In that study, the authors correlated gene expression with methylation data for *MOBP* and confirmed the association of the H1 haplotype with CpG methylation at the *MAPT* locus.

As in previous studies, a major limitation of our exploratory study is the lack of adequate statistical power because of the small sample sizes. In addition, regarding the issue of multiple comparisons (although *p* values for comparisons of each disease group with controls were adjusted using Dunnett's correction), we cannot exclude the possibility that our primary findings are false positives. However, the difference in CpG1 hypomethylation between PSP and controls was observed using an independent sample of 11 PSP brains and 12 controls as a replication cohort. Further and larger studies are warranted to confirm these preliminary results, which should be considered only as hypothesis-generating research.

We observed a mean difference of 4.1% in CpG1 hypomethylation in PSP-affected brains when compared with controls. Although low, this variation is consistent with results obtained in other studies on neurodegenerative diseases using postmortem brain tissues. Indeed, differences of less than 6% in methylation levels are commonly observed[15,21,23,29] and can be explained by a mixture of cells with and without methylation in brains.[15] Only a small percentage of the neurons and/or glial cells are affected in brain tissues presenting the hallmark of a neurodegenerative disease. Iwata and colleagues[15] used transcription activator-like effector constructs to confirm that even a small increase (less than 10%) in the methylation level can be associated with altered expression.

*MAPT* methylation levels were examined in 2 brain regions that are differently affected by PSP pathology: the frontal cortex, which contains pathological neurofibrillary tangles, and the occipital cortex, which is normally spared in PSP. Our results revealed hypomethylation in the PSP frontal cortex that was not observed in occipital tissues. This observed hypomethylation is therefore specific to the pathological tissue. However, differences between tissues are well known, especially within the human brain. Trabzuni and colleagues[30] reported an extensive study based on 2011 brain samples from 439 individuals and showed significant regional variations in *MAPT* mRNA expression and splicing. In a large study, Gibbs and colleagues[31] tested 27,578 DNA methylation sites and expression levels of 22,184 genes in 4 brain regions (the frontal cortex, temporal cortex, cerebellum, and pons) from 150 individuals. The authors correlated genetic variations, DNA methylation, and gene expression across the human genome (including variations in *MAPT*). However, many of these correlations differed across the 4 indicated brain regions. Regarding *MAPT* methylation, regional variations are present in AD as well in PD. Indeed, Iwata and colleagues[15] observed in the same AD patients differences in *MAPT* methylation between the temporal cortex, an area typically affected by neurofibrillary tangles, and the parietal cortex and the cerebellum, which are typically less affected.[12] Similarly, Coupland and colleagues[16] reported variations in *MAPT* methylation in the putamen and cerebellum, but not the anterior cingulate cortex, in PD subjects. These variations were correlated with the extent of PD pathology in these brain regions.

To investigate the effect of DNA methylation of intron 0, we assessed DNA methylation and *MAPT* mRNA expression in both of the brain regions we examined. We present the first report of overexpression of *MAPT* mRNA in PSP samples in frontal tissues, a phenomenon not observed in other tauopathies. However, we did not observe any correlation between DNA methylation of intron 0 and *MAPT* expression. We hypothesize that *MAPT* expression is regulated by other gene regions and likely depends on the use of a different alternative promoter. The use of these different promoters can be regulated by epigenetic modifications that generate different transcripts and/or interfere with splicing.[32] Thus, different tauopathies may involve different epigenetic modifications of *MAPT*, constituting specific epigenetic signatures.

Indeed, the promoter linked to the large CpGI in exon 0 might not be the unique promoter region of *MAPT*.[33,34] Analysis of *MAPT* transcripts from the literature and databases revealed several alternative TSSs within exon 0 or exon 1 of *MAPT* that might correspond to alternative promoters. More than half of all human genes are regulated by alternative promoters (APs), with an average of 3.1 putative APs per gene.[35] AP usage, which has previously been suggested in AD,[36,37] appears to be a major mechanism for determining the regional differences in gene expression during old age.[38] This regulation of APs is controlled by intragenic DNA methylation.[39,40] For example, DNA demethylation of intragenic CpGIs may result in use of the alternative downstream promoter via releasing methyl binding domains, facilitating transcription factor binding and/or DNA looping to a distal enhancer.[19] Moreover, AP regulation may also explain the variability in brain regions affected in tauopathies. Indeed, AP usage is highly tissue specific.[35,38,40] Notably, the brain is the tissue containing the second largest number of tissue-specific putative APs.[35] Thus, AP dysregulation of many genes, such as *MAPT*, may be involved in the physiopathology of a number of neurodegenerative diseases. Our study demonstrating an alteration of DNA methylation in PSP patients in an intragenic region of *MAPT* reinforces this hypothesis.

We present convergent data (hypomethylation, mRNA overexpression, disease, and region specificity) indicating that epigenetic changes may be involved in PSP. Altogether, our results emphasize the importance of this control region in intron 0 in PSP physiopathology. Our findings also demonstrate the involvement of multiple MAPT regions in tauopathies. Additional investigations of DNA and RNA levels using a larger cohort will be necessary to assess the roles of these epigenetic signatures in the transcriptional deregulation of MAPT in PSP and other neurodegenerative diseases.

**Acknowledgements:**

We acknowledge Amélie Labudeck for her technical contribution, Patrick Gelé as engineer in charge of the Lille brainbank and Sabrina Leclère as engineer in charge of the Paris brainbank. We thank Samuel Malone for his valuable contribution to the translation of this publication.

**REFERENCES:**

**1.** Boeve BF. Progressive supranuclear palsy. Parkinsonism Relat Disord 2012;18(suppl 1):S192-S194.

**2.** Williams DR, Lees AJ. Progressive supranuclear palsy: clinicopathological concepts and diagnostic challenges. Lancet Neurol 2009;8(3):270-279.

**3.** Golbe LI. Progressive supranuclear palsy. Semin Neurol 2014; 34(2):151-159.

**4.** Hauw JJ, Verny M, Delaere P, Cervera P, He Y, Duyckaerts C. Constant neurofibrillary changes in the neocortex in progressive supranuclear palsy. Basic differences with Alzheimer's disease and aging. Neurosci Lett 1990;119(2):182-186.

**5.** Hof PR, Delacourte A, Bouras C. Distribution of cortical neurofibrillary tangles in progressive supranuclear palsy: a quantitative analysis of six cases. Acta Neuropathol 1992;84(1):45-51.

**6.** Hauw JJ, Daniel SE, Dickson D, et al. Preliminary NINDS neuropathologic criteria for Steele-Richardson-Olszewski syndrome (progressive supranuclear palsy). Neurology 1994;44(11):2015-2019.

**7.** Hoglinger GU, Melhem NM, Dickson DW, et al. Identification of common variants influencing risk of the tauopathy progressive supranuclear palsy. Nat Genet 2011;43(7):699-705.

**8.** Baker M, Litvan I, Houlden H, et al. Association of an extended haplotype in the tau gene with progressive supranuclear palsy. Hum Mol Genet 1999;8(4):711-715.

**9.** Houlden H, Baker M, Morris HR, et al. Corticobasal degeneration and progressive supranuclear palsy share a common tau haplotype. Neurology 2001;56(12):1702-1706.

**10.** Pastor P, Moreno F, Clarimon J, et al. MAPT H1 haplotype is associated with late-onset Alzheimer's disease risk in APOEvarepsilon4 noncarriers: results from the Dementia Genetics Spanish Consortium. J Alzheimers Dis 2015;49(2):343-352.

**11.** Pittman AM, Myers AJ, Abou-Sleiman P, et al. Linkage disequilibrium fine mapping and haplotype association analysis of the tau gene in progressive supranuclear palsy and corticobasal degeneration. J Med Genet 2005;42(11):837-846.

**12.** Myers AJ, Pittman AM, Zhao AS, et al. The MAPT H1c risk haplotype is associated with increased expression of tau and especially of 4 repeat containing transcripts. Neurobiol Dis 2007;25(3):561-570.

**13.** Hernandez DG, Nalls MA, Gibbs JR, et al. Distinct DNA methylation changes highly correlated with chronological age in the human brain. Hum Mol Genet 2011;20(6):1164-1172.

**14.** Li Y, Chen JA, Sears RL, et al. An epigenetic signature in peripheral blood associated with the haplotype on 17q21.31, a risk factor for neurodegenerative tauopathy. PLoS Genet 2014;10(3): e1004211.

**15.** Iwata A, Nagata K, Hatsuta H, et al. Altered CpG methylation in sporadic Alzheimer's disease is associated with APP and MAPT dysregulation. Hum Mol Genet 2014;23(3):648-656.

**16.** Coupland KG, Mellick GD, Silburn PA, et al. DNA methylation of the MAPT gene in Parkinson's disease cohorts and modulation by vitamin E in vitro. Mov Disord 2014;29(13):1606-1614.

**17.** Barrachina M, Ferrer I. DNA methylation of Alzheimer disease and tauopathy-related genes in postmortem brain. J Neuropathol Exp Neurol 2009;68(8):880-891.

**18.** Yu CE, Cudaback E, Foraker J, et al. Epigenetic signature and enhancer activity of the human APOE gene. Hum Mol Genet 2013;22(24):5036-5047.

**19.** Kulis M, Queiros AC, Beekman R, Martin-Subero JI. Intragenic DNA methylation in transcriptional regulation, normal differentiation and cancer. Biochim Biophys Acta 2013;1829(11):1161-1174.

**20.** Tohgi H, Utsugisawa K, Nagane Y, Yoshimura M, Ukitsu M, Genda Y. The methylation status of cytosines in a tau gene promoter region alters with age to downregulate transcriptional activity in human cerebral cortex. Neurosci Lett 1999;275(2):89-92.

**21.** Sanchez-Mut JV, Aso E, Heyn H, et al. Promoter hypermethylation of the phosphatase DUSP22 mediates PKA-dependent TAU phosphorylation and CREB activation in Alzheimer's disease. Hippocampus 2014;24(4):363-368.

**22.** Pihlstrom L, Berge V, Rengmark A, Toft M. Parkinson's disease correlates with promoter methylation in the alpha-synuclein gene. Mov Disord 2015;30(4):577-580.

**23.** Chang L, Wang Y, Ji H, et al. Elevation of peripheral BDNF promoter methylation links to the risk of Alzheimer's disease. PLoS One 2014;9(11):e110773.

**24.** Caparros-Lefebvre D, Golbe LI, Deramecourt V, et al. A geographical cluster of progressive supranuclear palsy in northern France. Neurology 2015;85(15):1293-1300.

**25.** Horvath J, Burkhard PR, Bouras C, Kovari E. Etiologies of parkinsonism in a century-long autopsy-based cohort. Brain Pathol 2013; 23(1):28-33.

**26.** Koga S, Josephs KA, Ogaki K, et al. Cerebellar ataxia in progressive supranuclear palsy: an autopsy study of PSP-C. Mov Disord 2016;31(5):653-662.

**27.** Respondek G, Stamelou M, Kurz C, et al. The phenotypic spectrum of progressive supranuclear palsy: a retrospective multicenter study of 100 definite cases. Mov Disord 2014;29(14):1758-1766.

**28.** Allen M, Burgess JD, Ballard T, et al. Gene expression, methylation and neuropathology correlations at progressive supranuclear palsy risk loci. Acta Neuropathol 2016;132(2):197-211.

**29.** Foraker J, Millard SP, Leong L, et al. The APOE gene is differentially methylated in Alzheimer's disease. J Alzheimers Dis 2015; 48(3):745-755.

**30.** Trabzuni D, Wray S, Vandrovcova J, et al. MAPT expression and splicing is differentially regulated by brain region: relation to genotype and implication for tauopathies. Hum Mol Genet 2012; 21(18):4094-4103.

**31.** Gibbs JR, van der Brug MP, Hernandez DG, et al. Abundant quantitative trait loci exist for DNA methylation and gene expression in human brain. PLoS Genet 2010;6(5):e1000952.

**32.** Kornblihtt AR. Chromatin, transcript elongation and alternative splicing. Nat Struct Mol Biol 2006;13(1):5-7.

**33.** Caillet-Boudin ML, Buee L, Sergeant N, Lefebvre B. Regulation of human MAPT gene expression. Mol Neurodegener 2015;10:28.

**34.** Andreadis A. Tau gene alternative splicing: expression patterns, regulation and modulation of function in normal brain and neurodegenerative diseases. Biochim Biophys Acta 2005;1739(2-3):91-103.

**35.** Kimura K, Wakamatsu A, Suzuki Y, et al. Diversification of transcriptional modulation: large-scale identification and characterization of putative alternative promoters of human genes. Genome Res 2006;16(1):55-65.

**36.** Twine NA, Janitz K, Wilkins MR, Janitz M. Whole transcriptome sequencing reveals gene expression and splicing differences in brain regions affected by Alzheimer's disease. PLoS One 2011;6(1): e16266.

**37.** Drzewinska J, Walczak-Drzewiecka A, Ratajewski M. Identification and analysis of the promoter region of the human DHCR24 gene: involvement of DNA methylation and histone acetylation. Mol Biol Rep 2011;38(2):1091-1101.

**38.** Pardo LM, Rizzu P, Francescatto M, et al. Regional differences in gene expression and promoter usage in aged human brains. Neurobiol Aging 2013;34(7):1825-1836.

**39.** Maunakea AK, Nagarajan RP, Bilenky M, et al. Conserved role of intragenic DNA methylation in regulating alternative promoters. Nature 2010;466(7303):253-257.

**40.** Cheong J, Yamada Y, Yamashita R, et al. Diverse DNA methylation statuses at alternative promoters of human genes in various tissues. DNA Res 2006;13(4):155-167.

**AUTHOR CONTRIBUTIONS:**

V. Huin was involved in the research project (Conception, execution, analysis and interpretation of data), statistical analysis (Execution) and manuscript preparation (Writing of the first draft and revision of the manuscript)

V. Deramecourt was involved in in the research project (Analysis and interpretation of data of neuropathology) and manuscript preparation (Review and critique).

D. Caparros-Lefebvre was involved in in the research project (Analysis and interpretation of data of clinical history) and manuscript preparation (Revision of the manuscript).

C. A. Maurage was involved in in the research project (Analysis and interpretation of data of neuropathology).

C. Duyckaerts was involved in in the research project (Analysis and interpretation of data of neuropathology).

E. Kovari was involved in in in the research project (Analysis and interpretation of data of neuropathology).

F. Pasquier was involved in in in the research project (Analysis and interpretation of data of clinical history).

V. Buée-Scherrer was involved in in the research project (Interpretation of data of the brainbank).

J. Labreuche was involved in the statistical analysis (Design, execution, review and critique) and manuscript preparation (Writing of the first draft, review and critique).

H. Behal was involved in the statistical analysis (Design, execution, review and critique) and manuscript preparation (Writing of the first draft, review and critique).

L. Buée was involved in the study follow-up and manuscript preparation (Discussion and review and critique).

C. M. Dhaenens was involved in the research project (Conception, analysis and interpretation of data), and manuscript preparation (Writing of the first draft and review and critique).

B. Sablonnière was involved in the research project (Analysis, interpretation of data), and manuscript preparation (Review and critique).

**TABLES:**

|  | First cohort | | | | Replicate cohort | |
|---|---|---|---|---|---|---|
|  | Controls | AD | CBD | PSP | Controls | PSP |
| Subjects | 19 | 35 | 10 | 18 | 12 | 11 |
| Age of death (year-old) | 68.9±16.3 | 79.0±11.0 | 71.8±6.4 | 76.3±7.8 | 78.6±14.0 | 74.1±7.7 |
| PMI (h) | 15.1±9.2 | 15.7±12.2 | 19.8±14.1 | 19.9±11.3 | 25.9±12.2 | 15.8±11.1 |
| Sex Ratio (Men/Women) | 3.8 (15/4) | 0.8 (16/19) | 1.5 (6/4) | 1.3 (10/8) | 1 (6/6) | 1.2 (6/5) |
| Haplotype H1 of *MAPT*: *n* (%) | 30 (78.9) | 49 (70.0) | 13 (65.0) | 32 (88.9) | 21 (87.5) | 18 (81.8) |
| Diplotype H1/H1: *n* (%) | 11 (57.9) | 17 (48.6) | 6 (60.0) | 14 (77.8) | 8 (66.7) | 7 (63.6) |
| Diplotype H1/H2: *n* (%) | 8 (42.1) | 15 (42.8) | 2 (20.0) | 4 (22.2) | 3 (25.0) | 4 (36.4) |
| Diplotype H2/H2: *n* (%) | 0 (0) | 3 (8.6) | 2 (20.0) | 0 (0) | 0 (0) | 0 (0) |
| Allele A of rs242557: *n* (%) | 15 (39.5) | 26 (37.1) | 4 (22.2) | 15 (41.7) | 7 (35.0) | 11 (55.0) |

**Table 1:** Main characteristics of the postmortem cases analyzed in this study.

AD, Alzheimer's disease; CBD, corticobasal dementia; PSP, progessive supranuclear palsy; PMI, postmortem interval.

**FIGURE LEGENDS:**

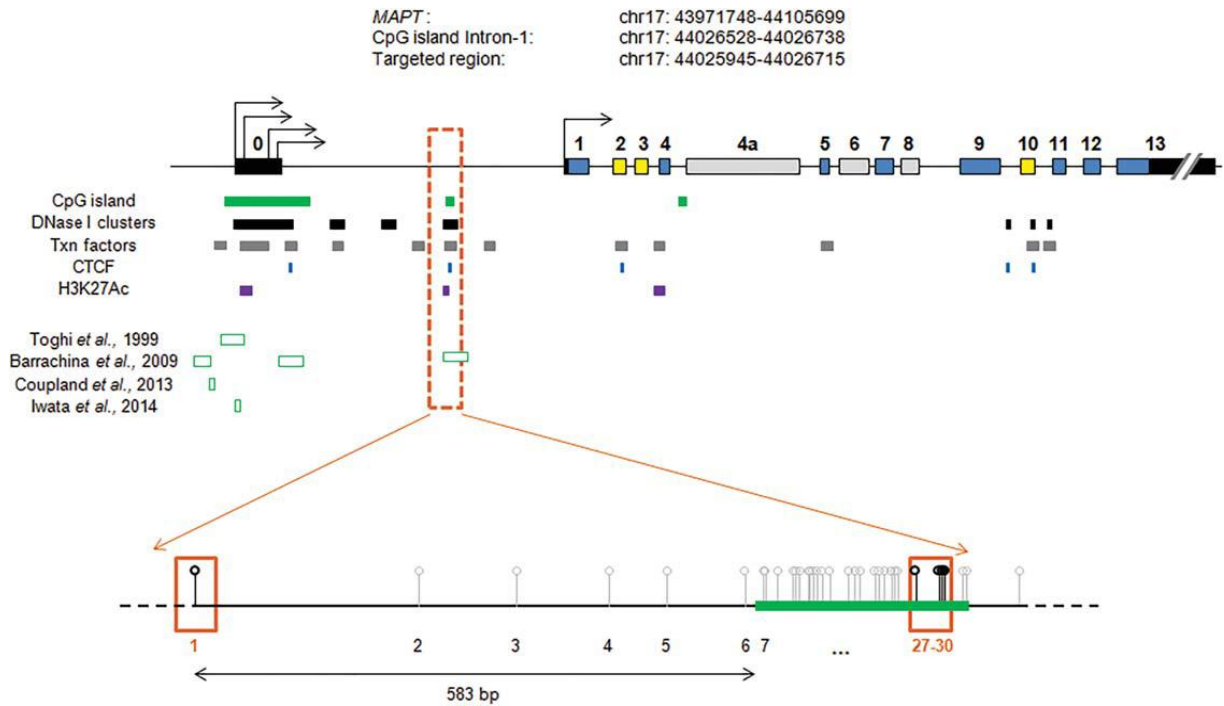

**Figure 1:** Schematic diagram of microtubule-associated protein tau (*MAPT*) gene and the position of the different CpG islands tested in the candidate region. Untranslated regions are indicated by black boxes, constitutive exons are marked in blue, alternative exons in the brain are marked in yellow, and exons presenting alternative splicing in other tissues are marked in gray. The different transcription initiation sites are depicted by black perpendicular arrows. Regulatory marks from University of California, Santa Cruz are shown in colored boxes under the gene. The different regions of *MAPT* assessed for DNA methylation in other studies are marked using empty boxes. Genomic regions analyzed in the studies of Iwata et al.,[15] Coupland et al.,[16] Barrachina et al.,[17] and Tohgi et al.[20] White lollipops represent CpGs in the targeted region in intron 0. The 5 CpGs analyzed were CpGs 1, 27, 28, 29, and 30. Txn: Transcription factor binding site; CTCF: CCCTC-binding factor binding site. Data are from Iwata et al.,[15] Coupland et al.,[16] Barrachina et al.,[17] and Tohgi et al.[20]

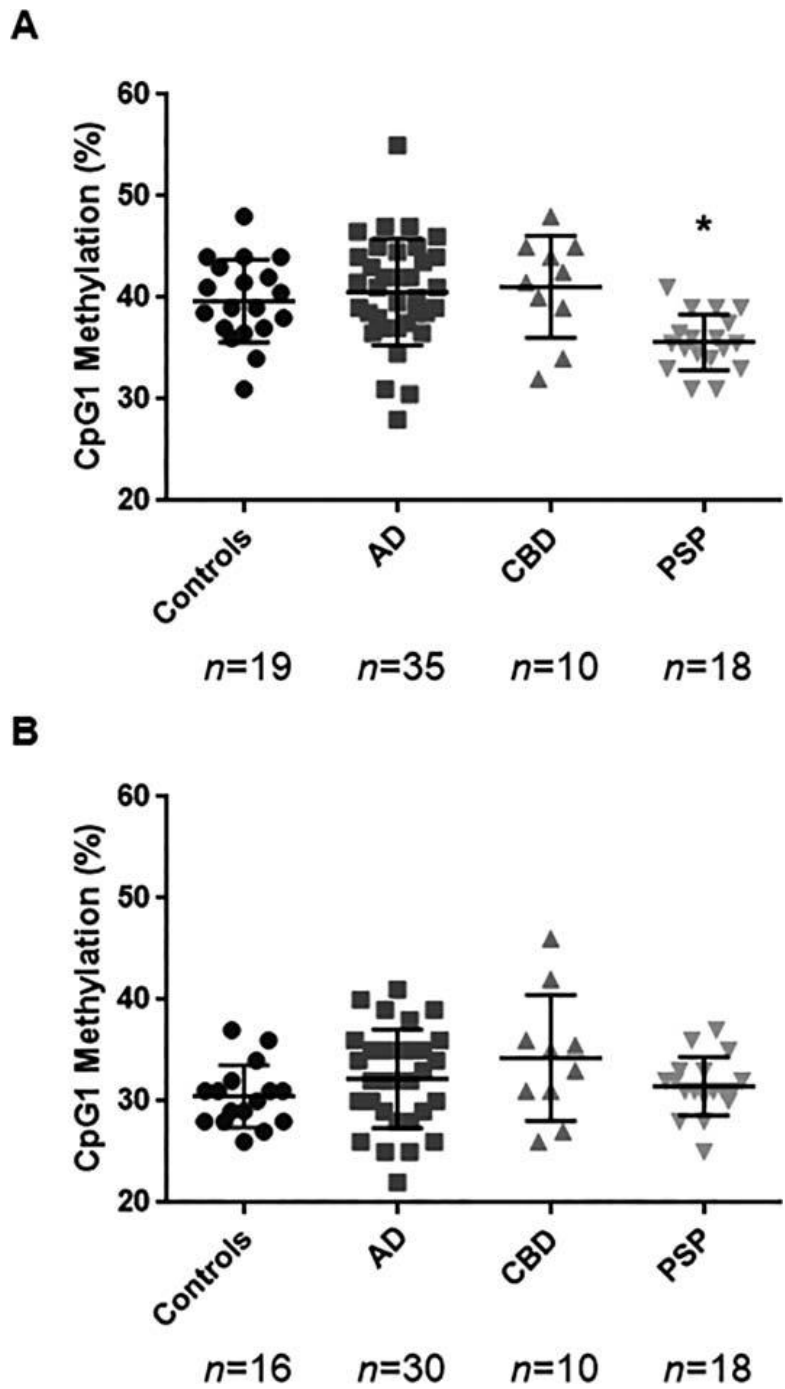

**Figure 2:** Comparison of *MAPT* CpG1 methylation levels in the frontal area (A) and in the occipital area (B) between tauopathies and controls. *$p<0.05$ for comparison with controls after adjustment for multiple comparisons (Dunnett's test). AD, Alzheimer's disease; CBD, corticobasal degeneration.

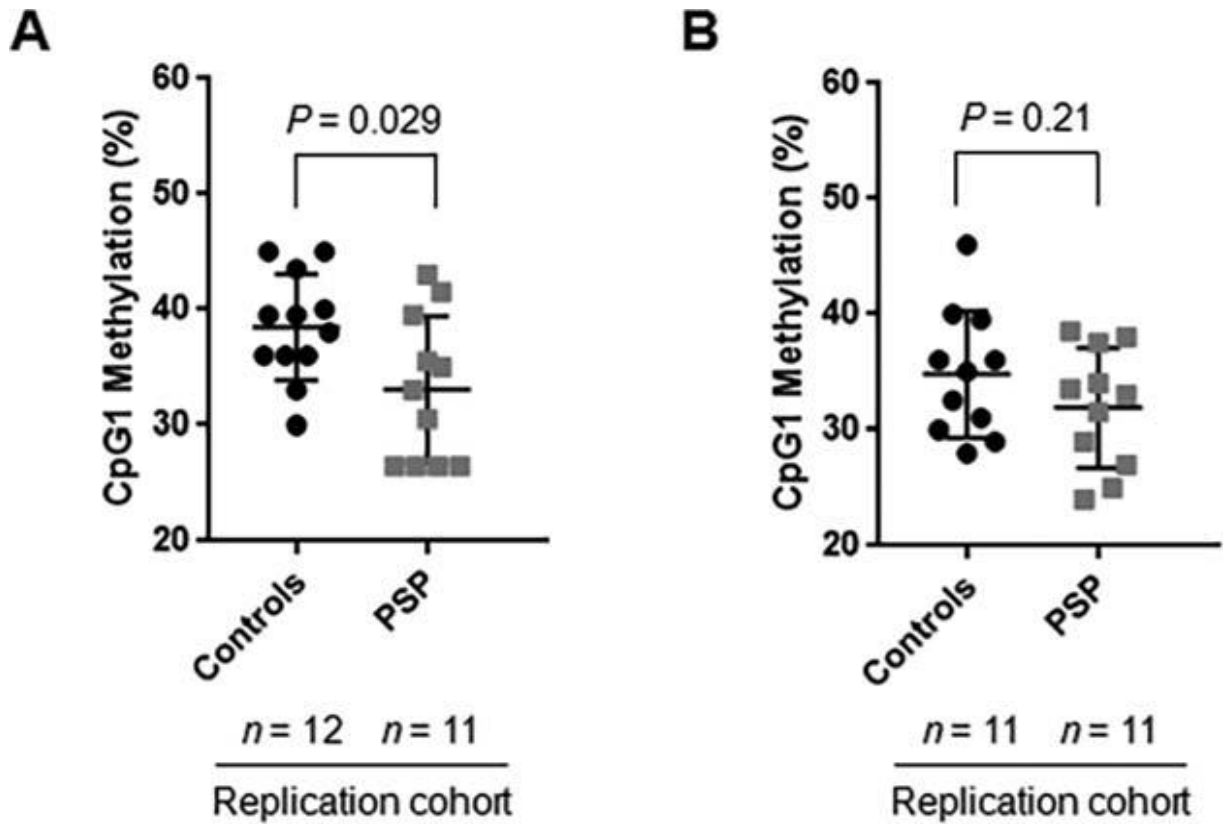

**Figure 3:** Replication analysis. Microtubule-associated protein tau gene CpG1 methylation levels in the frontal area (A) and in the occipital area (B) between PSP and controls. P values correspond to comparisons between controls and PSP samples using Student's *t* tests.

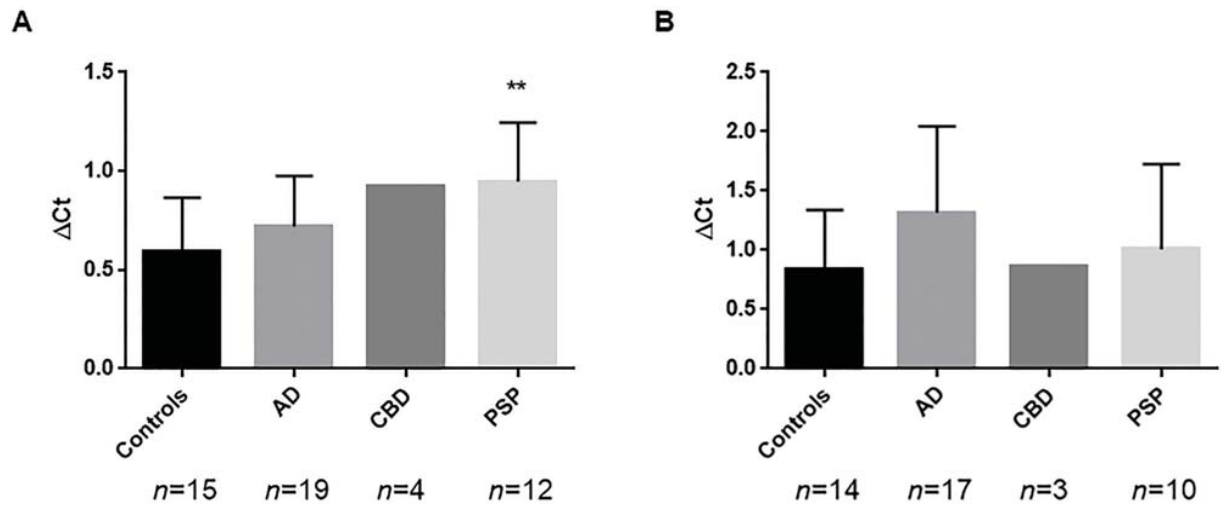

**Figure 4:** Microtubule-associated protein tau (*MAPT*) gene expression in postmortem brain based on the frontal area (A) and the occipital area (B) from the studied groups. The median values and interquartile ranges of the MAPT expression are reported for the different study groups. No statistical comparison between controls and the corticobasal degeneration (CBD) group was performed because of the small number of CBD samples with correct RNA integrity number. **$P < 0.01$ for comparison with controls (Mann–Whitney U test). AD, Alzheimer's disease.

**SUPPLEMENTAL MATERIALS:**

| Primer name | Size (bp) | Forward primer | Reverse primer | Sequencing primer |
|---|---|---|---|---|
| a. MAPT H1/H2 genotyping: | | | | |
| Int9-del-238 | 246-484 | 5'-GGAAGACGTTCTCACTGATCTG-3' | 5'-AGGAGTCTGGCTTCAGTCTCTC-3' | |
| b. MAPT H1c genotyping: | | | | |
| rs242557 | 293 | 5'-GAACATGCACATTTCTGCAAC-3' | 5'-AATGCTGGGAAGCAAAAGAA-3' | |
| c. Pyrosequencing: | | | | |
| Int-1.CpG1 | 155 | 5'-AAGGAGAAAGTTTTTTTAGGAAAT-3' | 5'-BIO-AAACTTCAAAACCCAAACATCC-3' | 5'-AGTTATTGTTTGATTTAATT-3' |
| Int-1.CpG34-37 | 181 | 5'-GGGATTTTTTTTGGTTGTTTTTTA-3' | 5'-BIO-AAAATCAAAAATCCAAAAACTCA-3' | 5'-GGATTTAGATTGGAAGT-3' |

**Supplemental Table S1:** List of primers.

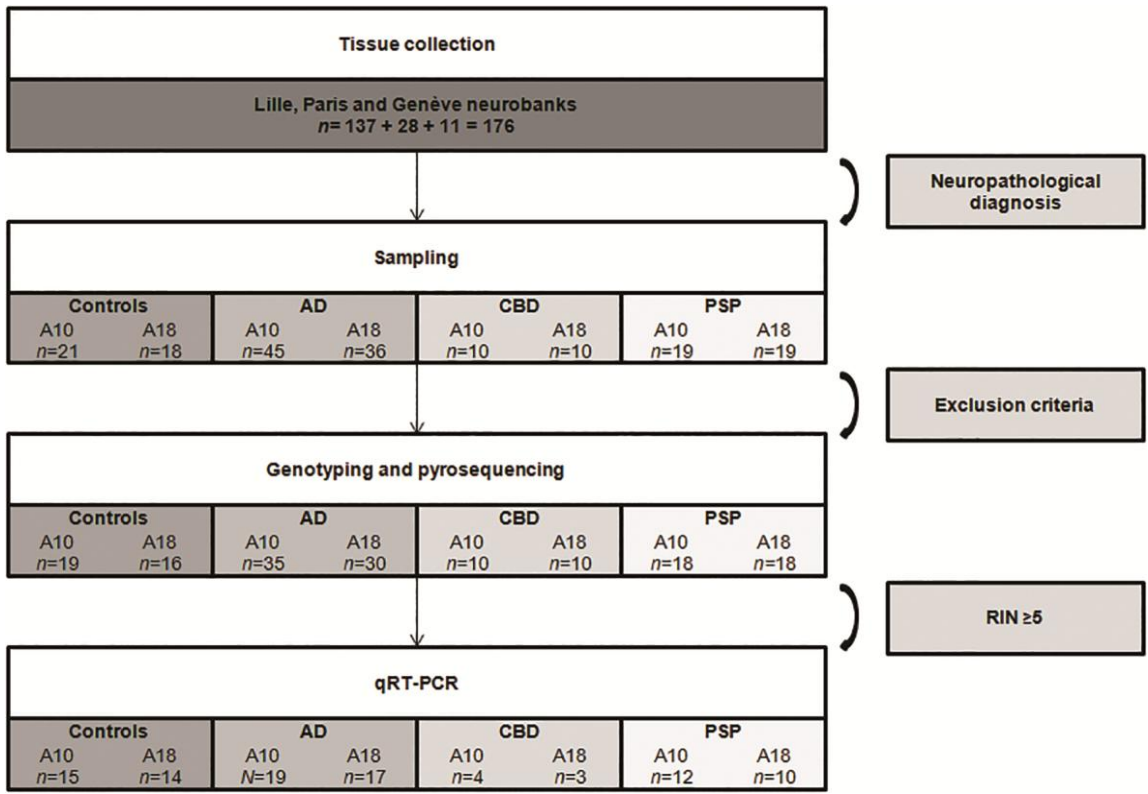

**Supplemental Figure S1:** Timeline of analyses on post-mortem brains. For each group, controls or patients, the number of samples tested in this study is indicated for the frontal samples (A10) and the occipital samples (A18). For the controls and the PSP group, the two numbers represent the respective numbers of samples from the first and second cohorts.

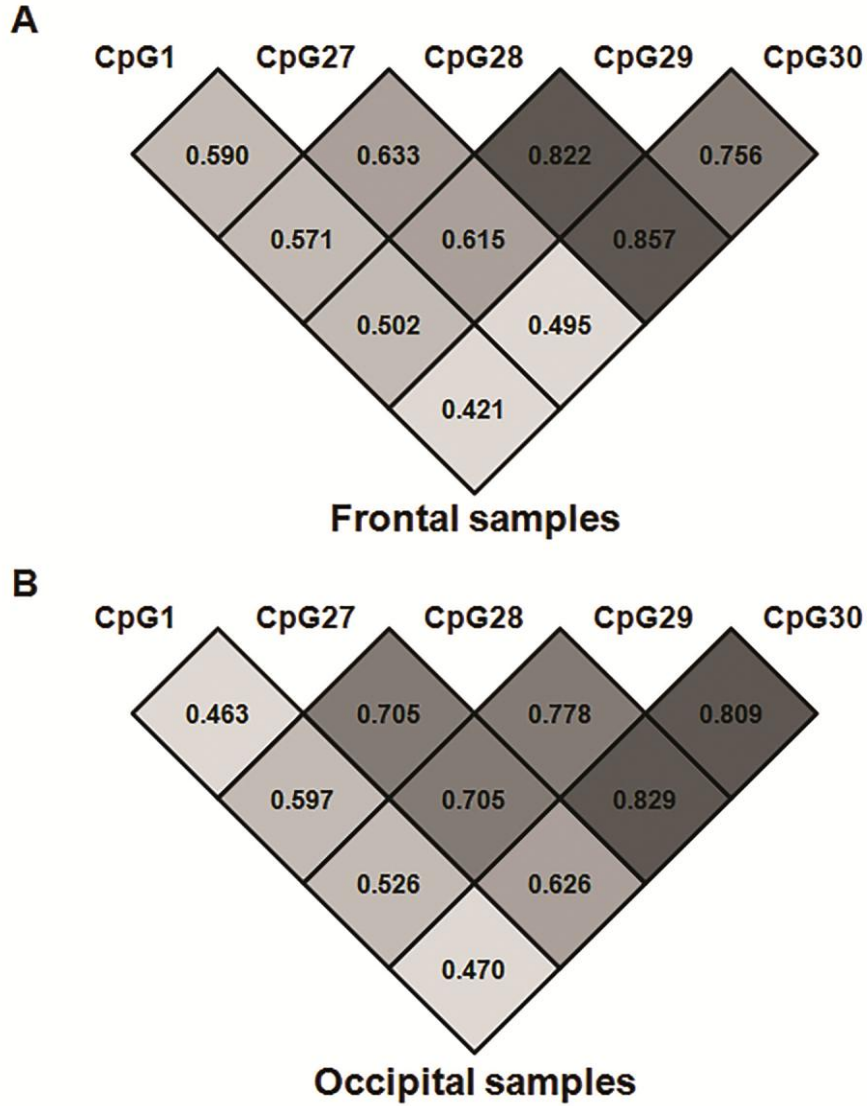

**Supplemental Figure S2:** Plots of the Pearson correlations of the methylation levels of the five CpGs tested. The figure shows the r values for each pair of CpGs and for the frontal samples, n=82 (A) and the occipital samples, n=77 (B). Note that methylation of all the CpGs had a significant positive correlation, but a stronger correlation was observed between CpG 28 and 30, with r>0.75. Numbers in the figure indicate r-values (P<0.001).

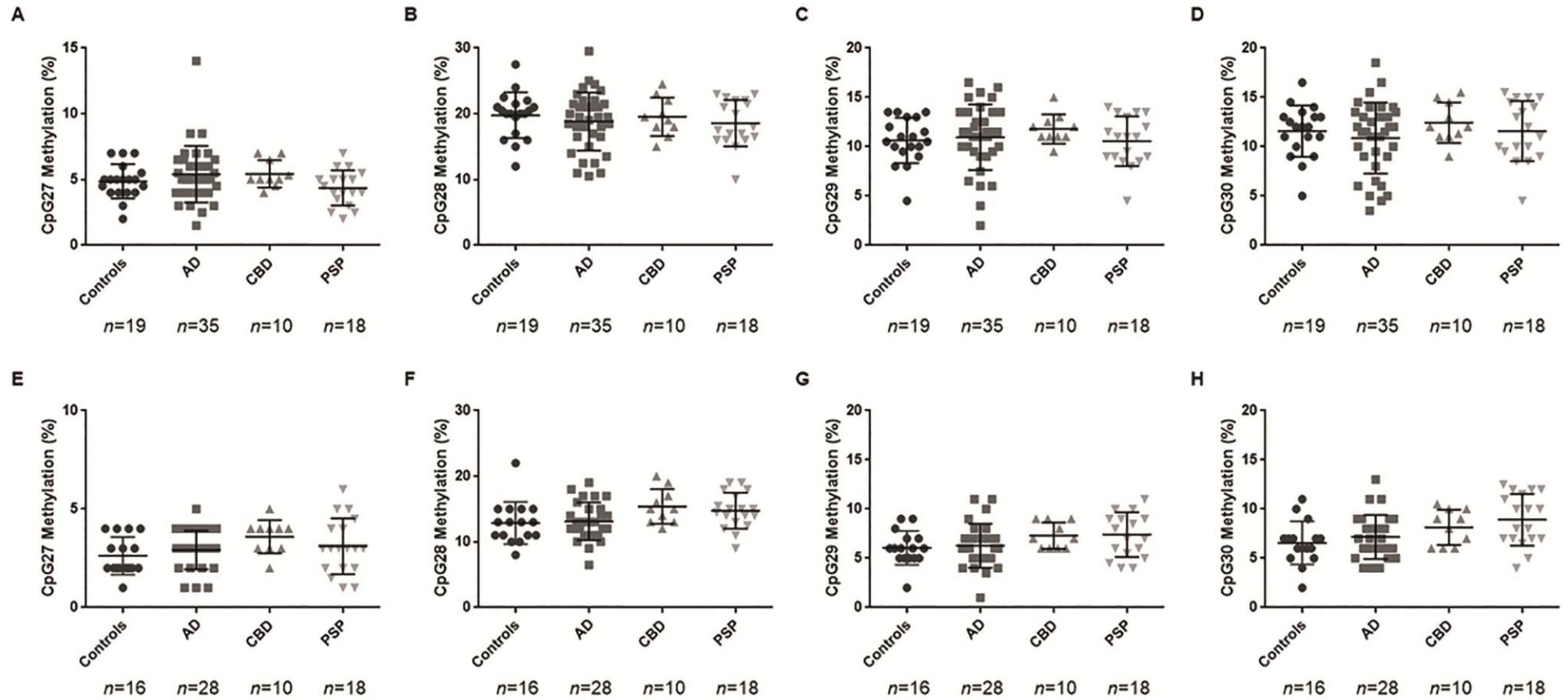

**Supplemental Figure S3:** Comparison of MAPT methylation levels in the frontal area (A) and in the occipital area (B) between tauopathies and controls.

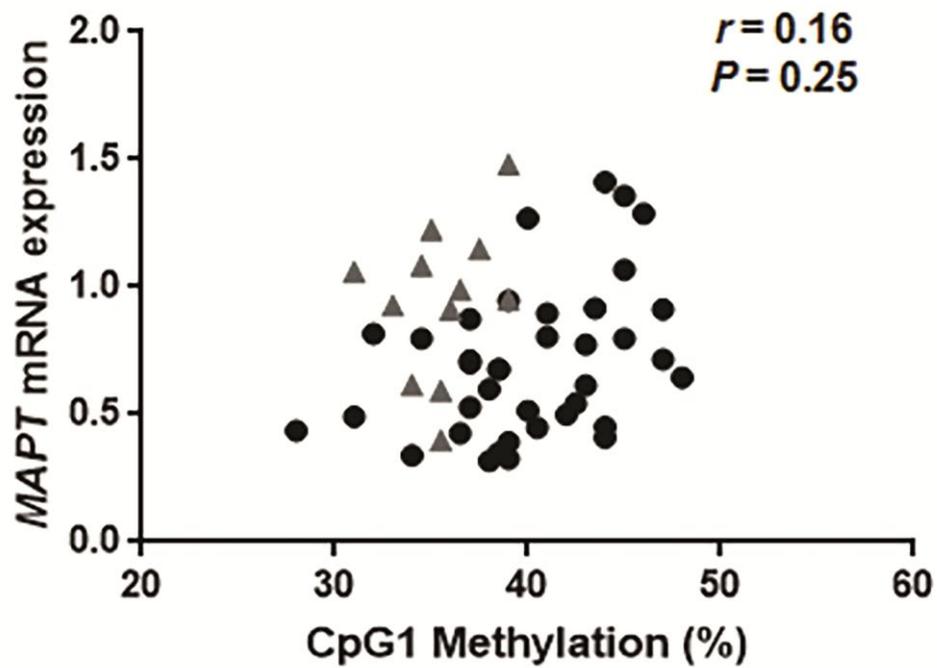

**Supplemental Figure S4:** CpG1 methylation levels and *MAPT* mRNA expression. The figure shows the correlation of CpG1 methylation levels and *MAPT* qRT-PCR results for the frontal samples (*n*=50). PSP samples are depicted by grey triangles.